\documentclass[pra,aps,twocolumn,superscriptaddress,notitlepage]{revtex4-2}
\usepackage{amssymb,amsfonts,amsmath,amstext,graphicx,hyperref,lipsum,empheq}
\usepackage{float}
\usepackage{placeins}

\usepackage[normalem]{ulem}
\hypersetup{
	colorlinks=true,
	linkcolor=blue,
	filecolor=magenta,      
	urlcolor=cyan,
}
\usepackage{amsthm}
\usepackage{outlines}

\newcommand{\tens}[1]{%
  \mathbin{\mathop{\otimes}\limits_{#1}}%
}

\usepackage{xparse}
\usepackage{outlines}
\usepackage{subfigure}
\usepackage{cleveref}
\usepackage{soul}

\usepackage[dvipsnames]{xcolor}

\newcommand{\beq}[0]{\begin{equation}}
\newcommand{\eeq}[0]{\end{equation}}
\newcommand{\non}{\nonumber}
\def\be{\begin{equation}}
\def\ee{\end{equation}}
\def\bea{\begin{eqnarray}}
\def\eea{\end{eqnarray}}
\newcommand{\ba}{\begin{eqnarray}}
\newcommand{\ea}{\end{eqnarray}}

\usepackage{hyperref}

\def\BraVert{\egroup\,\mid\,\bgroup}

\definecolor{myblue}{rgb}{.8, .8, 1}

\newcommand{\Lim}[1]{\raisebox{0.5ex}{\scalebox{0.8}{$\displaystyle \lim_{#1}\;$}}}

\begin{document}

\title{Stabilizing boundary time crystals through non-Markovian dynamics}
\author{Bandita Das}%
 \affiliation{Department of Physical Sciences, Indian Institute of Science Education and Research Berhampur, Berhampur 760010, India}
\author{Rahul Ghosh}%
\affiliation{Department of Physical Sciences, Indian Institute of Science Education and Research Berhampur, Berhampur 760010, India}
\author{Victor Mukherjee}%
\affiliation{Department of Physical Sciences, Indian Institute of Science Education and Research Berhampur, Berhampur 760010, India}

\date{\today}

\begin{abstract}
We study Boundary time crystals (BTCs) in the presence of non-Markovian dynamics. In contrast to BTCs observed in earlier works in the  Markovian regime, we show that non-Markovian dynamics can be highly beneficial for stabilizing BTCs  over a wide range of parameter values, even in the presence of intermediate rates of dissipation. Notably, we also observe the emergence of higher-order limit cycles (HO-LCs) for some parameter regimes. We analyze the effect of non-Markovian dynamics on BTCs and HO-LCs using quantum Fisher information, time-averaged magnetization, a measure of non-Markovianity, and a dynamical phase diagram, all of which show complex behaviors with changing non-Markovianity parameters. Our studies can pave the way for stabilizing time crystals in dissipative systems, as well as lead to studies on varied dissipative dynamics on time translational symmetry breaking.

\end{abstract}

\maketitle

\section{Introduction}
\label{secI}

The dynamics of many-body quantum systems driven out of equilibrium is a major field of theoretical \cite{dziarmaga10dynamics, dutta15quantum,delmontemeasurement2025,santinisemiclassical2025,correale2024finitefrequencyprethermalizationperiodicallydriven} and experimental \cite{king22coherent} research. Specifically, several open questions remain regarding thermalisation in driven quantum systems, which is in turn connected to a relatively newly discovered non-equilibrium phase of matter - viz. time crystals, which are formed in the presence of spontaneous time-translational symmetry breaking (TTSB). Discrete time crystals are formed in quantum systems driven periodically in time \cite{sacha2015modeling,zaletel2023colloqium,khemani2017defining, paltemporal2018} in the presence of disorder \cite{choi2017observation}, as well as in disorder-free many-body quantum systems \cite{kshetrimayum20stark}, for example, in the presence of long-range interactions \cite{floquetrussomanno2017,giachettifractal2023}. On the other hand, continuous or boundary time crystals (BTCs) are associated with limit cycles in open quantum systems in the presence of time-independent Hamiltonians \cite{iemini2018boundary,prazeres2021_boundary,kongkhambut22observation}. Interestingly, dissipative discrete time crystals have been shown to be related to quantum engines \cite{zaletel2023colloqium}, while BTCs can be a prime candidate for the development of many-body autonomous engines \cite{carallo2020engine}. In addition, use of time crystals to model qubits for quantum processing tasks has also been proposed \cite{autti21ac}. On the fundamental side, the thermodynamics of time crystals has received significant attention lately \cite{Carollo_2024Quantum}. Consequently, finding methods of generating and stabilizing time-crystals in varied many-body systems is a crucial question in this field. This becomes even more pertinent due to the fact that while weak dissipation has been shown to favour the existence of time crystals \cite{RieraCampeny2020timecrystallinityin,gong2018dtc,kessler2021observation,piccitto2021symmetries,sahoo2025generatingdiscretetimecrystals}, strong dissipation can have a detrimental effect on the same, for discrete time crystals \cite{gong2018dtc,Zhu_2019dicke,das2024discrete} as well as for BTCs \cite{ iemini2018boundary,solankiexotic2024,solanki2024chaostimedissipativecontinuous}. 

Most of the works on time crystals in open quantum systems have focused on Markovian dynamics. On the other hand,   non-Markovian dynamics, which may arise due to strong coupling or comparable dimensions of system and bath, may become an inescapable phenomenon in realistic systems \cite{breuer07the,Xia2024Markovian,shen2025emergentnonmarkoviangainopen}. Therefore, studying the fate of time crystals in the presence of non-Markovian dynamics is an important open question. Notably, the intriguing question remains if information backflow associated with non-Markovian dynamics can be harnessed to increase the stability of time crystals in the presence of strong dissipation.   Recently, it was shown that non-Markovian dynamics can be significantly beneficial for stabilizing discrete time crystals in the presence of intermediate rates of dissipation, in the semi-classical limit \cite{das2024discrete}. However, whether the benefits of non-Markovian dynamics extend to BTCs as well is an open question, which we address in this work. As we discuss below, non-Markovian dynamics can indeed allow the existence of BTCs  and higher order limit cycle (HO-LC) phases in the presence of intermediate rates of dissipation, as compared to Markovian dynamics, which can support BTCs only for weak rates of dissipation. We support our analysis using quantum Fisher information (QFI) and time-averaged magnetization, both of which show interesting characteristics associated with the BTC to non-BTC phase transition. In addition, we also use a measure of non-Markovianity $\mathcal{N}$ to show that intermediate values of $\mathcal{N}$ can be specially beneficial for the generation of BTCs and HO-LCs.  Finally, we summarize our results using a phase diagram which shows the behavior of BTCs and HO-LCs for different strengths of non-Markovianity and rates of dissipation.

We describe the model and dynamics of the dissipative setup considered here in Sec. \ref{secIIA},  focus on the non-Markovian regime in Sec. \ref{secIIB}, discuss QFI in Sec. \ref{secIIC}, study the behaviour of time-averaged magnetization in Sec. \ref{secIID}, analyze BTC and HO-LC phases with respect to the measure of non-Markovianity $\mathcal{N}$ in Sec. \ref{secIIE}, and show a complete phase diagram in Sec. \ref{secIIF}. Finally, we conclude in Sec. \ref{secIII}.

\section{Model and dynamics}
\label{secII}
%%%%%%%%%%%%%%%%%%%%%%%%%%%%%%%%%%%%%%%%%%%%%%%%%%%%%%%%%%%%%%%%%%%%%%%%%%%%%%%%%%%%%%%%%%%%%%%%%%%%%%%%%%%%%%%%%%%%%%%%%%%%%%%%%%%%%%%%%%%%%%%%%%%%%%%%%%%%%%%%%%%%%%%%%%%

\subsection{Dissipative Boundary time crystals}
\label{secIIA}
BTCs are associated with the continuous spontaneous breaking of time translational symmetry. 
We consider the setup introduced in Refs. \cite{iemini2018boundary,walls1978cooperative}; the system considered here comprises $N_b$ driven two-level spins, termed as the boundary, that are collectively coupled to a Bosonic field mode comprising the bulk of the setup. Under the Markovian approximation, the  system represented by the collective spin degrees of freedom evolves according to a time-independent Gorrini-Kossakowski-Lindblad-Sudarshan master equation, where coherent Hamiltonian driving competes with collective spin decay \cite{breuer07the}.
At sufficiently strong coherent driving, the magnetization exhibits oscillatory behaviour, with its lifetime increasing indefinitely as the number of spins grows. While finite-size systems eventually settle into a stationary equilibrium value, in the thermodynamic limit, these oscillations persist indefinitely, signaling the spontaneous breaking of continuous time-translation symmetry. Moreover, in the BTC phase, the collective dynamics form closed periodic orbits that characterise the system’s long-term evolution.\\

The many-body system is described by the time-independent Hamiltonian which is given by  \cite{iemini2018boundary}
\ba
H_b = \omega_0 \hat{S_x} + \frac{\omega_x}{S}(\hat{S}_x)^2 + \frac{\omega_z}{S}(\hat{S}_z)^2.
\label{hamiltonian}
\ea
Here $\hat{S}_{\mu}=\frac{1}{2}\sum_k \hat{\sigma}_{\mu}^{k}$ represents collective spin degrees of freedom,  $\hat{\sigma}_{\mu}^{k}(\mu=x,y,z)$ are the Pauli matrices acting on the $k$-th spin, and $\omega_0$ is the frequency of the two level system.
The Lindblad equation describing the evolution of the state $\hat{\rho_b}$ of the system is given  by \cite{breuer2016colloquim, zhang2012_general}
\begin{equation}
    \frac{d\hat{\rho_b}}{dt}= i[\hat{\rho_b},\hat{H_b}]+\frac{\kappa(t)}{S}(\hat{S}_{-}\hat{\rho}_b\hat{S}_+-\frac{1}{2}\bigl\{\hat{S}_{+}\hat{S}_{-},\rho_b \bigr\}).
    \label{mastereq}
\end{equation}
where we set $\hbar = 1$, $\hat{S}_{\pm}$ are the  raising and lowering operators i.e. $\hat{S}_{\pm} = \hat{S}_{x} \pm i  \hat{S}_{y} $ and we fix the total spin $S=N_b/2$. The time-dependent decay rate $\kappa(t)$ encodes the dissipative coupling to the environment.

The expectation value of an arbitrary operator $\hat{O}$ evolves following the open-system Heisenberg equation~\cite{gong2018dtc, iemini2018boundary}
\begin{align}
\frac{d}{dt} \langle \hat{O} \rangle 
= i\,  \!\left\langle [\hat{H}_b, \hat{O}] \right\rangle & + \frac{\kappa(t)}{2S}\, \!\left\langle [\hat{S}_+, \hat{O}] \, \hat{S}_{-}  + \hat{S}_+ [\hat{O}, \hat{S}_{-}]   \right\rangle
\label{eqO}
\end{align}
where $\langle \ldots \rangle = \mathrm{Tr}(\ldots \hat{\rho}) $. We introduce the normalized collective magnetization components
\begin{equation}
\hat{m}_\mu = \frac{\langle \hat{S}_\mu \rangle}{S}, \qquad \mu=x,y,z.
\label{eqavgm}
\end{equation}
From the spin commutation relations, one gets $\left[\hat{m}_\alpha, \hat{m}_{\beta} \right] = i \epsilon_{\alpha \beta \gamma} \hat{m}_{\gamma}/S$.
In the thermodynamic limit $N_b\to\infty$, one can neglect terms of the order of $\sim 1/S^2$, such that using Eqs. \eqref{eqO} and \eqref{eqavgm}, we arrive at a semi-classical mean-field dynamics described by the following equations~\cite{prazeres2021_boundary}:
\begin{subequations}
\label{eq:meanfield}
\begin{align}
\frac{d m_x}{dt} &= -2\omega_z\, m_y m_z + \kappa(t)\, m_x m_z, \\
\frac{d m_y}{dt} &= 2(\omega_z - \omega_x)\, m_x m_z - \omega_0\, m_z + \kappa(t)\, m_y m_z, \\
\frac{d m_z}{dt} &= \omega_0\, m_y - \kappa(t)\left( m_x^2 + m_y^2 \right) + 2\omega_x\, m_x m_y .
\end{align}
\end{subequations}
The above equations couple all three components of the Bloch vector and conserve the total magnetization norm,
\begin{equation}
\mathcal{M} = m_x^2 + m_y^2 + m_z^2 = 1.
\end{equation}
 We consider $\omega_x=\omega_z=0$ in the main text, while the case of non-zero $\omega_x, \omega_z$ is discussed in Appendix \ref{appA}.

 To study the effect of non-Markovian dynamics on the behaviour of BTC, we consider a phenomenological model analogous to the damped Jaynes-Cummings model describing the decay of a two-level system coupled to Bosonic modes with a Lorentzian spectral function (see Ref. \cite{breuer07the} and Appendix \ref{appB} for more details). To this end, we assume  $\kappa(t)$ of the form

\begin{equation}
{ \kappa (t) } =  \begin{cases}
\frac{2 \gamma \kappa_0\sinh(t\omega_d)}{2 \omega_d \cosh(t\omega_d)+ \gamma \sinh (t\omega_d)} & \text{$|\kappa|< \kappa_{max}$}\\
\kappa_{max} & \text{ $|\kappa|\geq \kappa_{max}$} ,
\end{cases}  
\label{eqkappa}
\end{equation}
where $\gamma$ denotes the spectral width  of the coupling to the bath, while $\kappa_0$ is related to the Markovian system-bath coupling strength, and $\omega_d = (\sqrt{\gamma^2-2\gamma \kappa_0})/2$.  $\kappa_{max} > \kappa_0$ is a parameter which can be tuned to control the maximum possible rate of dissipation and, while in the case of the damped Jaynes-Cummings model, $\kappa_{max} \to \infty$. The above form of $\kappa(t)$ (Eq. \eqref{eqkappa}) allows us to tune between the Markovian  regime  ($\gamma>2\kappa_0$) in which case $\kappa(t) > 0~ $for all $~t$ and $\kappa(t\to\infty)=\frac{2\kappa_0 \gamma}{2 \omega_d +\gamma}$, and the non-Markovian regime ($\gamma < 2\kappa_0$), in which case $\kappa(t)$ assumes an oscillatory form realized by replacing $\sinh$ ($\cosh$) by $\sin$ ($\cos$) in Eq. \eqref{eqkappa}, and  can take negative values for some time intervals \cite{mukherjee2015efficiency}. Here complete positivity is ensured by $\int^t_0 \kappa(t) dt \geq 0$ for all $t$ \cite{cruzenski2014degree,breuer2009measure}.
 
We emphasise that despite of the phenomenological form of $\kappa(t)$, the time-local form of the master equation \eqref{mastereq} and the ability  for continuous tuning between the Markovian and the non-Markovian regime make this model an ideal platform for understanding the role of non-Markovian dynamics and information backflow on the fate of BTCs; furthermore, the Jaynes-Cummings model (cf. Eq. \eqref{eqkappa}) has also been realised experimentally in optical cavity setups \cite{rempe87Observation}.

\begin{figure}
    \centering
    \includegraphics[width=0.8\linewidth]{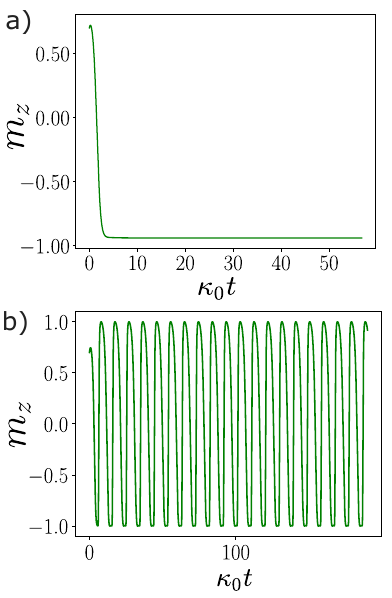}
    \caption{ Plot showing the dynamics of $m_z$ in the (a) Markovian ($\gamma=3\kappa_0$ $\forall~t$) and the (b) non-Markovian ($\gamma = \kappa_0/4$) regimes, for $\omega_0/\kappa_0 = 0.3$. As shown here, the Markovian regime results in a time-independent steady state, while the non-Markovian regime is associated with a BTC phase.
    }

    \label{fig:comparison}
\end{figure}

%\subsubsection{Markovian regime} 
BTCs were realised in the above model \eqref{mastereq} for a Markovian dissipative environment with a time-independent rate of dissipation ($\gamma \gg \kappa_0$; $\kappa(t) = \kappa_0 > 0$ $\forall~t$),    where persistent oscillations were shown by the $m_z$ in the thermodynamic limit, for small $\kappa_0$ ($\frac{\omega_0}{\kappa_0}>1$) \cite{iemini2018boundary}. On the other hand, stronger rates of dissipation ($\frac{\omega_0}{\kappa_0}<1$) result in replacement of the BTC phase by a time-independent steady state (TISS) (see Fig. \ref{fig:comparison}a). In the next section, we tune the value of $\gamma/\kappa_0$ from $\gamma \gg \kappa_0$ to $\gamma < \kappa_0$, to study the behaviour of the system in both the Markovian ($\gamma > 2\kappa_0$), as well as the non-Markovian ($\gamma < 2\kappa_0$) regime. We have set $\kappa_0=1$ and $\kappa_{max}=5$ for all the numerical results shown here.

\subsection{Non-Markovian regime}
\label{secIIB}

In this section we focus on the non-Markovian regime ($\gamma < 2 \kappa_0$), characterized by a  $\kappa(t)$ assuming negative values for some time intervals (see Eq. \eqref{eqkappa}), and associated with the so called information back-flow  \cite{bylicka17constructive}. In contrast to the behavior reported for Markovian dynamics \cite{iemini2018boundary} where BTC phase is present only for  $\frac{\omega_0}{\kappa_0}>1$, numerical analysis shows that  non-Markovian dynamics makes makes BTCs significantly more robust against lower values of $\omega_0$, as signified by the presence of time-crystalline order for $\frac{\omega_0}{\kappa_0}<1$ (see Figs.~\ref{fig:comparison} - \ref{fig:phasediagram}). This robustness of the BTC  phase w.r.t. $\omega_0/\kappa_0$ suggests the beneficial role played by information backflow in stabilizing time-crystalline order in the presence of intermediates rates of dissipation \cite{breuer2009measure, hsieh19non}. We study the behavior for different values of $\omega_0$ in Fig. \ref{fig:btcnon-Markovian}.  Small values of $\omega_0/\kappa_0$ are associated with a non-BTC phase, characterized by irregular oscillations of $m_z$ w.r.t. time $t$ (see Fig. \ref{fig:btcnon-Markovian}a), which correspond to spurious peaks in the corresponding FFT for different frequencies $\omega$ (see Fig. \ref{fig:btcnon-Markovian}d), and irregular trajectories in the Bloch sphere (see Fig. \ref{fig:btcnon-Markovian}g). As we increase $\omega_0/\kappa_0$, we obtain a BTC phase for intermediate values of $\omega_0/\kappa_0$, as shown by persistent oscillations in $m_z$ (see Fig. \ref{fig:btcnon-Markovian}b), one dominant and other less dominant peaks in the fast fourier transform (FFT) (see Fig. \ref{fig:btcnon-Markovian}e), and a limit cycle on the Bloch sphere  (see Fig. \ref{fig:btcnon-Markovian}  h). Interestingly, as we increase $\omega_0/\kappa_0$ further, we get HO-LCs  (see Fig. \ref{fig:btcnon-Markovian}c), supported by multiple dominant Fourier peaks (Fig.~\ref{fig:btcnon-Markovian}f)) and a limit cycle on the Bloch sphere (Fig.~\ref{fig:btcnon-Markovian}i)). 
Numerical analysis suggests the system remains in the HO-LC phase with varying strengths of the FFT peaks even for larger values of $\omega_0/\kappa_0$, with the long-time dynamics converging to a limit cycle. 
\begin{widetext}

\begin{figure}
    \centering
    \includegraphics[width=0.9\textwidth]{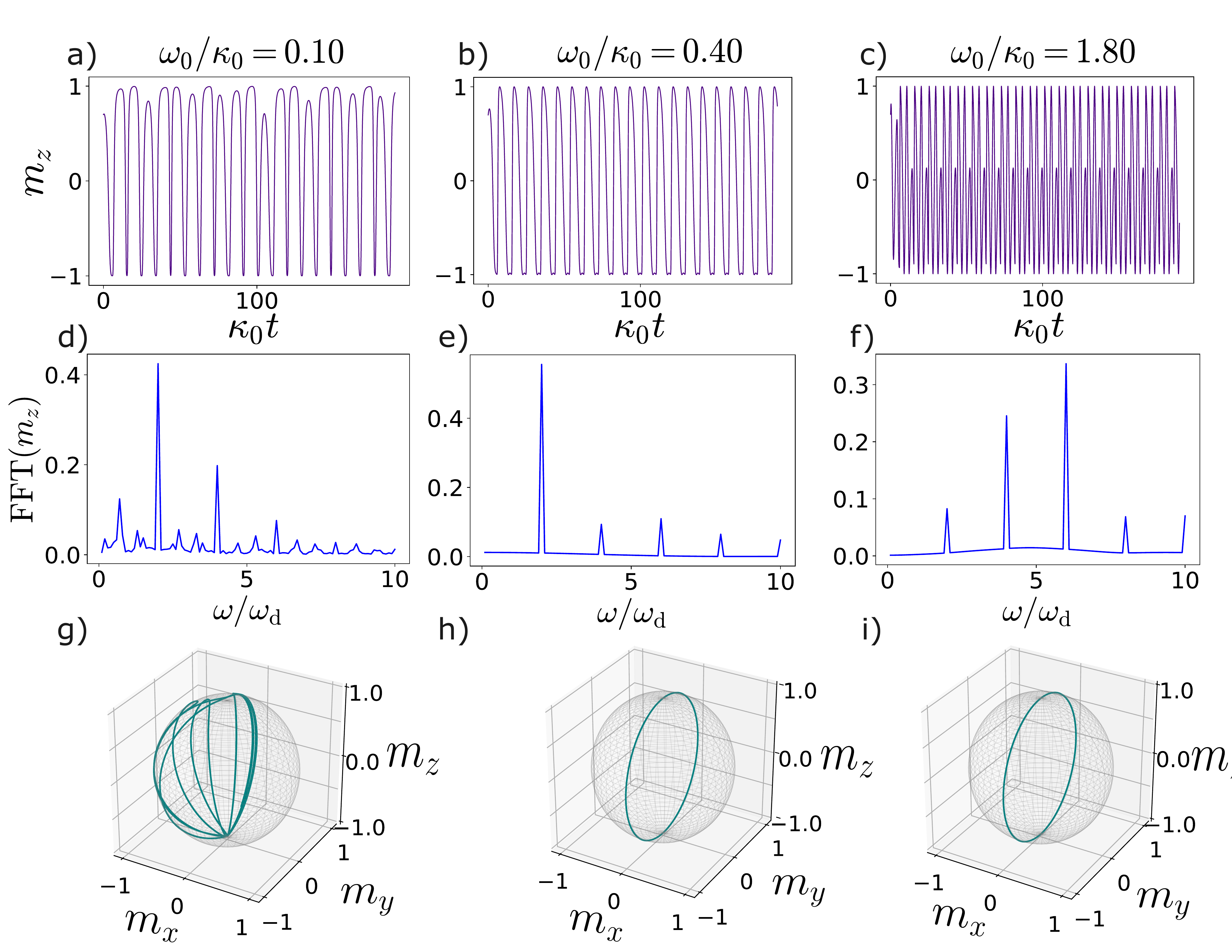 }
    \caption{Plot showing the dynamics of $m_z$ in the non-Markovian regime for $\gamma=\kappa_0/4$, (a, b, c), the corresponding FFT (d, e, f) and the representation of the dynamics on the Bloch sphere (g, h, i), for different values of $\omega_0/\kappa_0$, in the non-Markovian regime. $m_z$ shows (a) irregular oscillations with (d) multiple spurious peaks in the FFT and (g) an irregular trajectory on the Bloch sphere for small $\omega_0/\kappa_0$ . A moderate value of $\omega_0/\kappa_0$ is associated with a BTC phase, characterized by (b) an oscillatory $m_z$ with (e) a dominant frequency and multiple less dominant frequencies in the FFT and (h) a limit cycle on the Bloch sphere. A higher value of $\omega_0/\kappa_0$ results in (c) $m_z$ showing oscillations with (f) multiple dominant frequencies and (i) a limit cycle on the Bloch sphere.  }
  
    \label{fig:btcnon-Markovian}
\end{figure}

Unless otherwise stated,  here we consider  $(m_{x}^0,m_{y}^0,m_{z}^0) = (-0.58,\,0.417,\,0.7)$ as the initial state in all the numerical results presented here. However, as shown in Appendix \ref{appC}, the long time dynamics in time-translational symmetry broken regime (BTC and HO-LC) is robust to the choice of the initial state, while the same is not true in the non-BTC phase.
    
\end{widetext}

\subsection{Quantum Fisher Information}
\label{secIIC}
In this section, we analyse the transition between the BTC and non-BTC phases using quantum Fisher information (QFI). QFI has been widely used for studying the sensitivity of quantum probes \cite{iemini2024floquet,dominic2025boundary}, and also for detecting quantum phase transitions through divergences in QFI \cite{montenegro23quantum}. Here we compute the QFI for various values of $\omega_0$, in order to check the BTC to non-BTC transition in this model. In order to calculate the QFI, we first evaluate the density matrix within the mean-field approximation. Following  Ref.~\cite{dominic2025boundary}, we write the total density matrix $\hat{\rho}_b = \hat{\rho}_{\text{MF}, \omega_0}$, for $N_b$ two-level systems, in a factorized form as
\ba 
\hat{\rho}_{\text{MF},\omega_0} = \bigotimes_{i=1}^{N_b} \hat{\rho}_{i,\omega_0} ,
\label{rhomf}
\ea
where $\hat{\rho}_{i,\omega_0}$ denotes the reduced density matrix for the $i$-th spin.  These reduced states are identical for all spins and can be represented in terms of the total spin observables as \cite{Nielsen_Chuang_2010}
\ba
\hat{\rho}_{i,\omega_0} = \frac{1}{2}(\hat{\mathbb{I}}_2 + 
m_x\hat{\sigma}^x + m_y\hat{\sigma}^y + m_z\hat{\sigma}^z )
\label{rhomean}
\ea
where $\hat{\mathbb{I}}_2$ is the Identity matrix.
After taking into account that the system is in a separable state, and $\mathcal{F} (\hat{\rho}_A \tens{} \hat{\rho}_B ) = \mathcal{F} (\hat{\rho}_A) + \mathcal{F} (\hat{\rho}_B )$ for any two states $\hat{\rho}_A$ and $\hat{\rho}_B$, we get the total QFI by adding up the QFI $\mathcal{F}(\hat{\rho}_{i,\omega_0})$ of the reduced state of each qubit, such that~\cite{dominic2025boundary,montenegro23quantum}
\begin{align} \label{qfieq}
\mathcal{F}(\hat{\rho}_{\text{MF},\omega_0}) &= N_b \mathcal{F}(\hat{\rho}_{i,\omega_0}) \nonumber\\
 &= N_b \left(8  \Lim{\delta \omega_0 \to 0} \frac{\Big( 1 -  \mathbb{F}( \hat{\rho}_{i, \omega_0 - \delta \omega_0}, \hat{\rho}_{i, \omega_0 + \delta \omega_0}) \Big)}{(2 \delta \omega_0)^2} \right). 
\end{align} 
Here  $\mathbb{F} ( \hat{\rho}_1, \hat{\rho}_2) =\operatorname{Tr} \left[ \sqrt{ \sqrt{\hat{\rho}_1} \, \hat{\rho}_2 
\sqrt{\hat{\rho}_1} } \right]$ denotes the fidelity between two density matrices $\hat{\rho}_1$ and $\hat{\rho}_2$, and $\delta \omega_0$ denotes an infinitesimal change in $\omega_0$.

In general phase transitions are associated with divergences in QFI, owing to large changes in the state of a system for a small change in the Hamiltonian parameters. As shown in Fig. \ref{fig:qfi}a, the behavior of the QFI suggests a non-BTC to BTC phase transition at $\omega_0/\kappa_0 \approx 0.15$, thus showing a significant advantage as compared to Markovian dynamics, where the corresponding phase transition  occurs at $\omega_0/\kappa_0 \approx 1$ for $\gamma \gg \kappa_0$ (see Ref. \cite{iemini2018boundary}). The multiple peaks for $\omega_0/\kappa_0 \lessapprox 0.15$  can be associated with the irregular dynamics shown in Fig. ~\ref{fig:btcnon-Markovian}a. In contrast, the QFI remains close to zero for larger  $\omega_0/\kappa_0$, suggesting absence of any phase transition beween the BTC and HO-LC phases (see Fig. \ref{fig:btcnon-Markovian}).  For very small $\omega_0/\kappa_0$, the dissipation dominates the dynamics, and takes the system to a non-BTC oscillatory state, characterized by a non-limit cycle trajectory on the Bloch sphere (see Appendix \ref{appD}).

\subsection{Time-averaged magnetization}
\label{secIID}

\begin{figure}[H]
        \centering
        \includegraphics[width=0.9\linewidth]{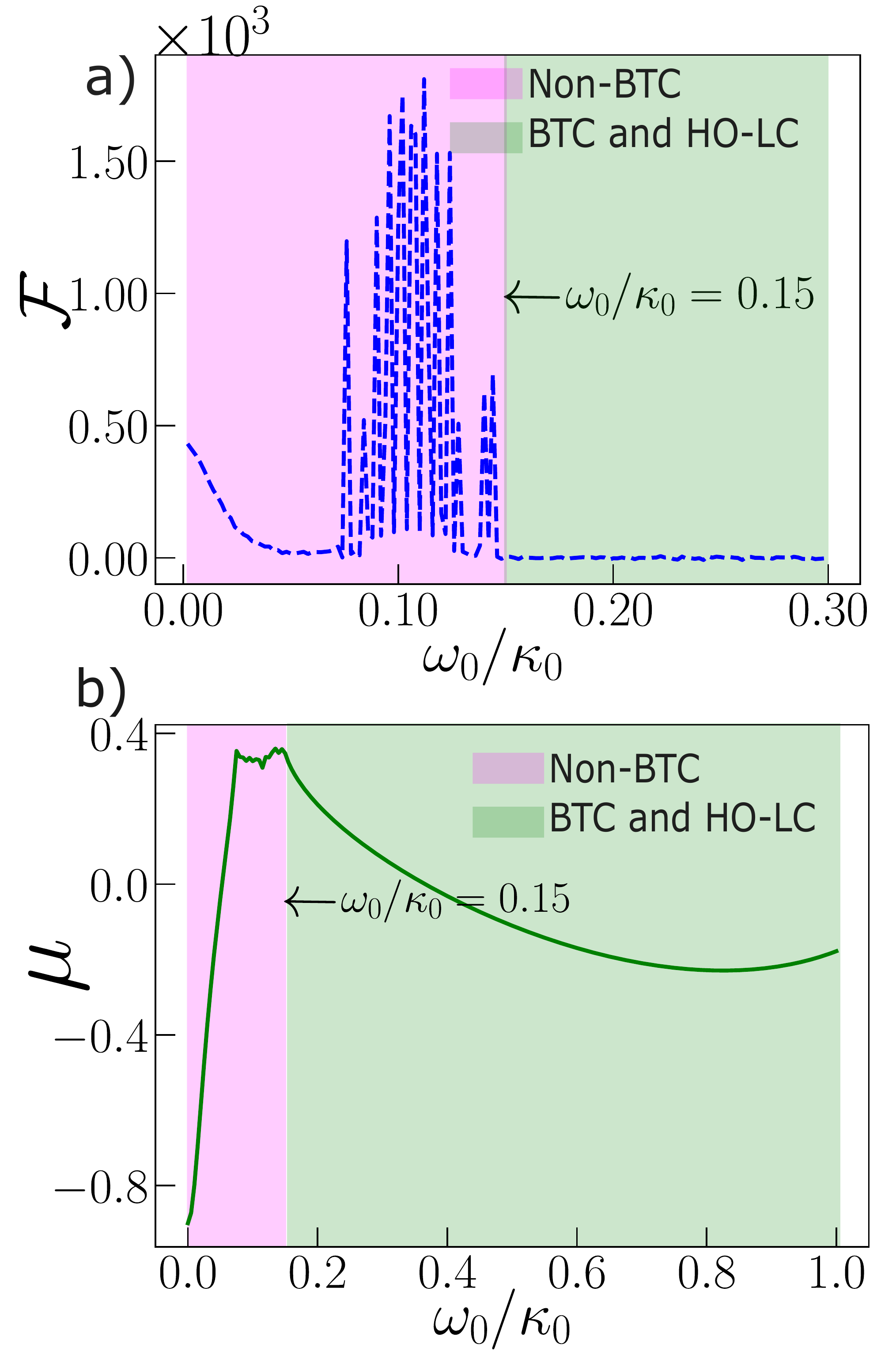}
        \caption{(a) We plot the QFI in the non-Markovian regime for $\gamma=\kappa_0/4$, as a function of $\omega_0/\kappa_0$. The transition between the BTC phase and the non-BTC phase is associated with divergences in QFI for $\omega_0/\kappa_0 \lessapprox 0.15$. (b) We plot the average magnetization $\mu$ as a function of $\omega_0/\kappa_0$. BTC to non-BTC transition at $\omega_0/\kappa_0$ is associated with distinct changes in the behavior of $\mu$.}
        \label{fig:qfi}
    
\end{figure}

To gain additional insight, we introduce the
time-averaged magnetization \cite{carollo2022exact,mattes2023_entangled},
\begin{equation}
\mu (\omega_0) = \frac{1}{t} \int_{0}^{t} m_z(\omega_0, t^{\prime})\, dt^{\prime}
\label{orderparam}
\end{equation}
as a diagonistic tool to study the  different dynamical regimes. We emphasize that, due to its time-averaged nature, $\mu$ may not be sensitive
to the presence of  oscillatory dynamics.
However, as presented in Fig. \ref{fig:qfi}b, numerical analysis shows that $\mu$ is associated with distinct behaviors in the non-BTC and in the time-translational symmetry broken phases (BTC and HO-LC) of this model.

In Ref. \cite{carollo2022exact} the time-averaged magnetization exhibits
distinct scaling behaviors across steady-state and oscillatory regimes and
is used as an order parameter to characterize BTC phase transition. While in our system the average
magnetization remains nonzero in all phases, $\mu$ nevertheless displays
qualitatively different behavior in the BTC / HO-LC and non-BTC regimes, separated
by a transition near $\omega_0/\kappa_0 \approx 0.15$, as shown in
Fig.~\ref{fig:qfi}b.  However, $\mu$ fails to distinguish between the BTC and the HO-LC phases, which can be attributed to the absence of any phase transition between these two phases, as shown by Fig. \ref{fig:qfi}a.

In the non-BTC regime, $m_z(t)$ exhibits oscillations that are biased
towards negative values for $\omega_0/\kappa_0 \rightarrow 0$
(see Appendix~\ref{appD}), resulting in a
negative value of $\mu$. For small but finite $\omega_0/\kappa_0$, the
irregular dynamics of $m_z(t)$
(Fig.~\ref{fig:btcnon-Markovian}a) leads to a correspondingly irregular but
predominantly positive behavior of $\mu$. In contrast, symmetric
oscillations in the BTC regime give rise to small values of $\mu$.

As the system crosses into the higher-order limit-cycle (HO-LC) regime,
the oscillations develop increasing asymmetry, with kink-like features
appearing in the magnetization dynamics
(Fig.~\ref{fig:btcnon-Markovian}c). This results in a systematic deviation
of $\mu$ away from zero, reflecting the increasing complexity of the
dynamics.

 \subsection{Non-Markovian Measure}
 \label{secIIE}
 One crucial question in the field of time crystals is, whether a clear relation exists between the existence of a time-crystalline phase and the degree of non-Markovianity present in the dynamics. To this end, we quantify the non-Markovianity using a decay-rate-based measure introduced in Refs. \cite{hall2014canonical,Rivas2014quantum}. Specifically, we define a function 
 \begin{equation}
   f(t)=max(-\kappa(t),0).  
 \end{equation}
This function is identically zero for all times if and only if the system's evolution is Markovian ($\kappa(t) \geq 0~\forall~t$); on the other hand a negative $\kappa(t)$ results in $f(t) = |\kappa(t)|$. Hence, any deviation from zero directly signals the presence of non-Markovian dynamics. A cumulative dimensionless measure of non-Markovianity can thus be constructed via the time integral \cite{Rivas2014quantum}
\ba
\mathcal{N} =\int_0^{t} f(t^{\prime}) dt^{\prime}.
\ea
A nonzero value of $\mathcal{N}$ implies that the decay rate $\kappa(t)$ becomes negative during certain intervals of time, signaling a temporary reversal of information flow from the environment back to the system \cite{breuer2016colloquim,piilo2008non-markovian}. 

\begin{figure}
    \centering
    \includegraphics[width=0.9\linewidth]{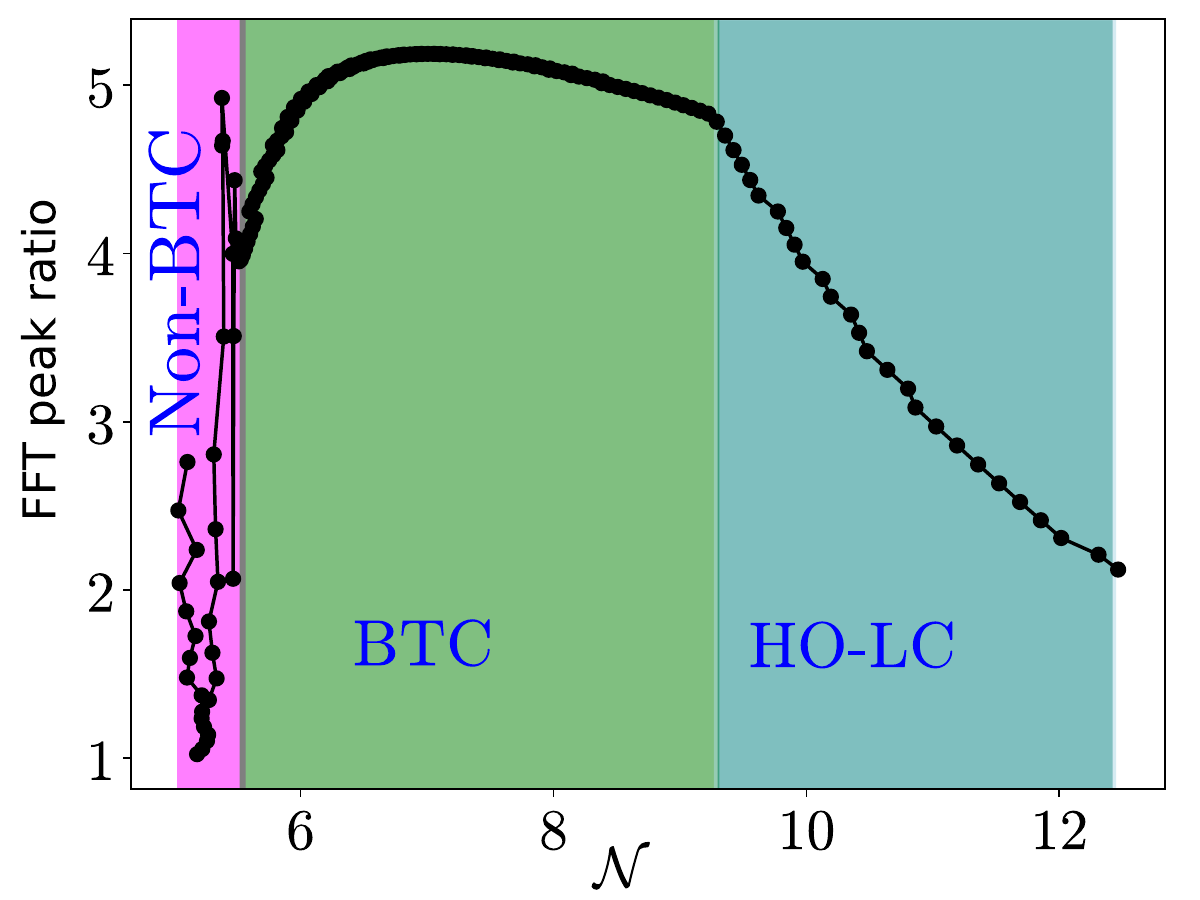}
    \caption{FFT peak ratio as a function of $\mathcal{N}$ for one period of $\kappa(t)$,  for a constant $\frac{\omega_0}{\kappa_0} = 0.55$. 
For small $\mathcal{N}$, the system stays in the non-BTC phase, and the FFT peak ratio fluctuates rapidly. In the BTC phase the ratio is finite and varies slowly, whereas in the HO--LC phase, it decreases due to the presence of multiple comparable dominant frequencies.  }
    \label{fig:peakratio}
\end{figure}

To characterize the dynamical phases, we consider the ratio between the amplitude of the dominant frequency peak and the amplitude of the second-largest peak in the FFT of $m_z$, here termed as the FFT peak ratio. Higher values of FFT peak ratio indicates the presence of a BTC phase, while lower values correspond to HO-LC or irregular non-BTC phases (see also Fig. \ref{fig:btcnon-Markovian}). 
 We plot the FFT peak ratio as a function of the non-Markovianity measure $\mathcal{N}$ in Fig. \ref{fig:peakratio},  which is varied by changing  $\gamma$ (see also Fig. \ref{fig:phasediagram}). For small values of $\mathcal{N}$, the peak ratio exhibits significant fluctuations, indicating an irregular non-BTC phase. As $\mathcal{N}$ increases with decreasing $\gamma$, the peak ratio reaches a maximum and varies slowly within the BTC regime. Beyond this region, as the system transitions into the higher-order limit cycle (HO-LC) regime, the peak ratio gradually decreases, reflecting the emergence of more complex dynamical structures with multiple competing frequencies. Notably, the FFT peak ratio is always finite, signifying the presence of multiple frequencies in the dynamics in the BTC and the HO-LC phases. The above results emphasize the importance of information-backflow for the generation of limit cycles, in the form of BTCs for intermediate values of $\mathcal{N}$, and as HO-LCs for higher values of $\mathcal{N}$.

\subsection{Phase diagram}
\label{secIIF}
\begin{figure}[h]
    \centering
     \includegraphics[width=1.1\linewidth]{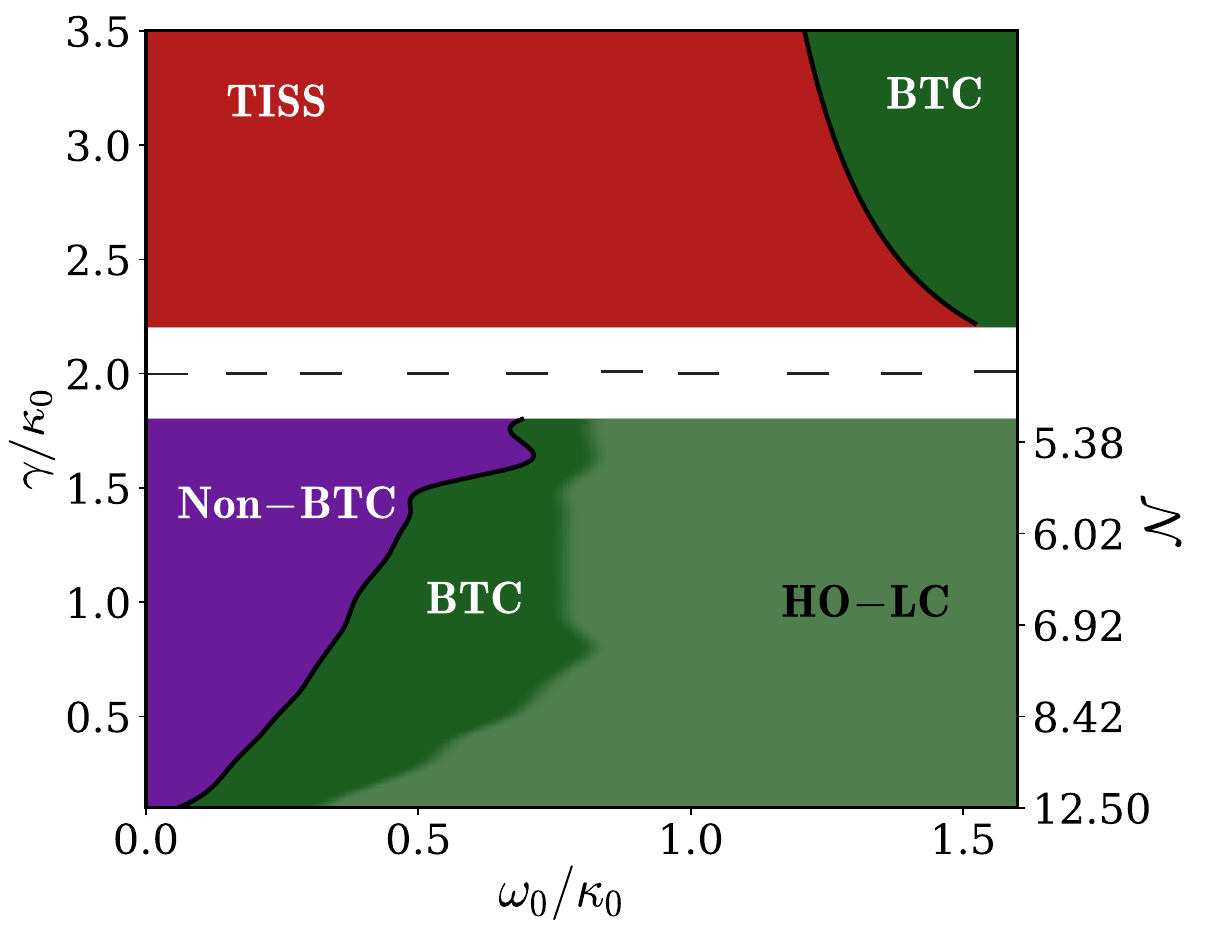}
    \caption{Phase diagram as a function of $\gamma/\kappa_0$ and $\omega_0/\kappa_0$, showing the crossover from the Markovian to the non-Markovian regime. The transition between the Markovian and non-Markovian regimes is indicated by the dashed line at $\gamma=2\kappa_0$. The  blank space denotes parameters close to $\gamma = 2\kappa_0$ where the time-scales diverge. The transition from the TISS or the non-BTC to the BTC phase is shown by the black solid lines. The corresponding $\mathcal{N}$ values in the non-Markovian regime for time $t = 10 \omega_d^{-1}$ is shown on the right.  }
    \label{fig:phasediagram}
\end{figure}
The qualitative behaviors of the system across the Markovian and non-Markovian  regimes are summarized in the phase diagram shown in Fig.~\ref{fig:phasediagram}. We  traverse between the Markovian regime ($\gamma  > 2 \kappa_0 $) and the non-Markovian regime ($\gamma  <2 \kappa_0 $) by tuning the parameter $\gamma$. This transition is clearly reflected in the system's dynamical response, as illustrated by the stark contrast observed on either side of $\gamma = 2\kappa_0$ (shown by the black dashed line in Fig.~\ref{fig:phasediagram}).

The time scales diverge close to $\gamma = 2\kappa_0$, shown by the blank space around the Markovian to non-Markovian transition line. The transition from the non-BTC phase to the BTC phase is  denoted by the black solid line, whereas the BTC and the HO--LC phases are not separated by any phase transition, as shown in Fig. \ref{fig:qfi}.

In the Markovian regime the system tends toward a time independent steady state in the limit of $\kappa(t \to \infty)$, here the time crystalline behaviour is suppressed by strong, memoryless dissipation. On the other hand, a BTC phase appears for $\omega_0 \gtrapprox \kappa(t \to \infty)$. Upon crossing into the non-Markovian regime, the memory effects become significant; here we observe the emergence of BTC, signified by persistent, robust oscillations in $m_z$, even for smaller values of $\omega_0/\kappa_0$. Further, in contrast to the Markovian regime, non-Markovian dynamics is characterized by a irregular  non-BTC phase for small  values of $\omega_0/\kappa_0$. On the other hand higher values of $\omega_0/\kappa_0$ are associated with HO-LC phase, for non-Markovian dynamics.

As shown in Fig.~\ref{fig:phasediagram}, the value of $\mathcal{N}$ increases monotonically as the system is driven deeper into the non-Markovian regime, characterized by smaller values of $\gamma$, while it remains zero in the Markovian regime ($\gamma > 2\kappa_0$). Interestingly, increase in $\mathcal{N}$ results in the BTC and HO-LC regimes occurring for smaller values of $\omega_0/\kappa_0$, indicating the beneficial effects of information backflow for making time crystals more robust against dissipation.

\section{Conclusion}
\label{secIII}

In this work we have shown that non-Markovian dynamics can be significantly beneficial for preserving BTCs in the presence of stronger rates of dissipation, in comparison to Markovian dynamics. 
In the Markovian regime, BTC phase is destroyed and gets replaced by a time-independent steady state for $\omega_0 / \kappa_0 < 1$; in contrast, as shown in Figs. \ref{fig:comparison} and \ref{fig:btcnon-Markovian}, information backflow for non-Markovian dynamics can preserve time crystalline order even for $\omega_0 / \kappa_0 < 1$. On the contrary, too small $\omega_0/\kappa_0$ results in a irregular  phase, while larger values of $\omega_0/\kappa_0$ give rise to higher-order limit cycle dynamical phase with multiple peaks in the FFT. In order to have a deeper understanding of the dynamical phases, we have studied the QFI as a function of $\omega_0/\kappa_0$. Interestingly, QFI does not show any signature of HO-LC to BTC transition, which can be attributed to the absence of any phase transition between these two phases, as also signified by the limit cycle feature for both the phases. On the other hand, transition between BTC to irregular  phase for small $\omega_0/\kappa_0$  is accompanied by multiple peaks in the QFI. The same phase transition is also manifested in the behavior of the time-averaged magnetization $\mu$, which rises with decreasing $\omega_0/\kappa_0$ in the ordered (BTC and HO-LC) phase, giving way to a irregular  regime  at the BTC - irregular  phase transition point,  finally showing a sharp decrease for small values of $\omega / \kappa_0$ (see Fig. \ref{fig:qfi}). 

In order to understand the relation between the different dynamical phases and information backflow associated with non-Markovian dynamics, next we have considered the non-Markovianity parameter $\mathcal{N}$; one can increase $\mathcal{N}$ by decreasing $\gamma/\kappa_0$, i.e., by entering deeper inside the non-Markovian regime, as shown in Fig. \ref{fig:phasediagram}. We have plotted the FFT peak ratio as a function of $\mathcal{N}$ in Fig. \ref{fig:peakratio}. The FFT peak ratio shows irregular  behavior and rises sharply with $\mathcal{N}$,      for small $\mathcal{N}$ in the irregular  regime, which changes to a smooth variation with $\mathcal{N}$ in the BTC phase (corresponding to intermediate values of $\mathcal{N}$), which changes to a sharp decrease with $\mathcal{N}$ in the HO-LC phase (corresponding to larger values of $\mathcal{N}$). This indicates the benefits of information backflow associated with non-Markovian dynamics for preserving BTC and HO-LC, and also raises questions regarding the relation between dynamical phases and other measures of non-Markovianity, such as quantum relative entropy \cite{goswami21experimental}.

Next we have plotted the phase diagram as a function of $\omega_0/\kappa_0$ and $\gamma/\kappa_0$ in Fig. \ref{fig:phasediagram}. As observed also for the discrete time crystal \cite{das2024discrete}, the Markovian to non-Markovian transition at $\gamma = 2\kappa_0$ is associated with distinct change in the phase diagram. In the Markovian phase we get BTC phase only for large $\omega_0/\kappa_0$, while smaller values of $\omega_0/\kappa_0$ are associated with a  TISS phase. In contrast, the phase diagram shows much richer behavior in the non-Markovian regime; irregular non-BTC  phase for small $\omega_0/\kappa_0$ gets replaced by a BTC phase for intermediate values of $\omega_0/\kappa_0$, which further changes to HO-LC phase for larger values of $\omega_0/\kappa_0$. Interestingly, the transition points between the above three regimes shift towards smaller values of $\omega_0/\kappa_0$ for smaller values of $\gamma/\kappa_0$, indicating small $\gamma/\kappa_0$ to be appropriate for getting the ordered phases (BTC and HO-LC) phases, if we are restricted to smaller ranges of $\omega_0/\kappa_0$. 

 The presence (absence) of robustness of the long-time dynamics to the choice of the initial state in the BTC / HO-LC (non-BTC) phase (see Appendix \ref{appC}) suggests one can use this dynamical feature to define an order parameter, in analogy with Ref. \cite{giachettifractal2023}; such an order parameter can be expected to take distinctly different values in the non-BTC and BTC / HO-LC phases. However, a detailed analysis of the above form of order parameter is beyond the scope of the present work.

We emphasize that although the dynamics in the BTC  and HO-LC phases are oscillatory and arise in the presence of a time-dependent $\kappa(t)$, the underlying mechanism differs fundamentally from that of discrete or Floquet time crystals. In Floquet time crystals the time-period of the response is in general integral \cite{zaletel2023colloqium} or even non-integral \cite{pawel19fractional} multiples of the period of the time-dependent  system Hamiltonian. In contrast, here the form of the phenomenological master equation (2) can arise in the presence of time-independent system and bath Hamiltonians, with the periodicity of $\kappa(t)$ arising inherently from the bath spectral function  (see Appendix \ref{appB}). Consequently, the dynamics of the response is not determined by any external driving frequency of the system Hamiltonian; rather, it depends on the details of the bare system and bath parameters, as is the case for BTCs [14]. Further,  the time-translational symmetry broken phases (BTC and HO-LC) are associated with multiple peaks in FFT (see Fig. \ref{fig:btcnon-Markovian}), in analogy with that observed in a BTC  in the Markovian limit \cite{iemini2018boundary}. Numerical analysis suggests that the frequencies of these peaks depend on $\omega_d$, while their amplitudes vary  with $\gamma/\kappa_0$ and  $\omega_0/\kappa_0$; the variation of the FFT peak ratio with the non-Markovianity parameter $\mathcal{N}$ is shown in Fig. \ref{fig:peakratio}.  We plan to study the detailed dependence of the FFT peaks on the system and bath parameters ($\gamma,  \kappa_0, \omega_0$) in future works. 
Furthermore, the fate of BTCs and HO-LCs in the presence of other forms of decay rates $\kappa(t)$, for example, arising in the presence of Ohmic, sub-Ohmic or super-Ohmic bath spectral functions, might be interesting as well. Information backflow and non-zero values of $\mathcal{N}$ is a signature of generic non-Markovian dynamics. Consequently, we expect time-crystalline order to be more robust in the presence of the other forms of $\kappa(t)$ with large enough $\mathcal{N}$ as well.  However, further rigorous analysis is needed to have a more definitive picture regarding the behavior of time crystals in the presence of generic non-Markovian dynamics.

Finally, we note that the studies considered here can be realized experimentally using already available technologies; non-Markovian dynamics along with measures of non-Markovianity has been studied experimentally  using quantum optical setups \cite{rempe87Observation} and machine learning models \cite{goswami21experimental}. Recently BTC has been realized in a dissipative atom-cavity setup \cite{kongkhambut22observation}, comprising $^{87}{\rm Rb}$ atoms in a high-finesse optical cavity. In this setup, the frequencies of the order of $\sim$ kHz were considered.  Therefore we envisage similar setups with decay rates $\kappa_0 \sim $ kHz and $\gamma < 2\kappa_0$ might allow us to experimentally realize robust time crystalline phases in the presence of strong dissipation; this in turn can help us to understand more about time translational symmetry breaking, and also show us ways to make BTCs and HO-LCs more robust to dissipation.

\section*{ACKNOWLEDGEMENTS}
B.D. acknowledges support from Prime Minister Research Fellowship(PMRF).  V.M. acknowledges support from Science and Engineering Research Board (SERB) through MATRICS (Project No.
MTR/2021/000055).

\appendix \label{app}

\section{Non-zero $\omega_x$ and $\omega_z$}
\label{appA}

\begin{figure}[h]
     \centering
     \includegraphics[width=0.8\linewidth]{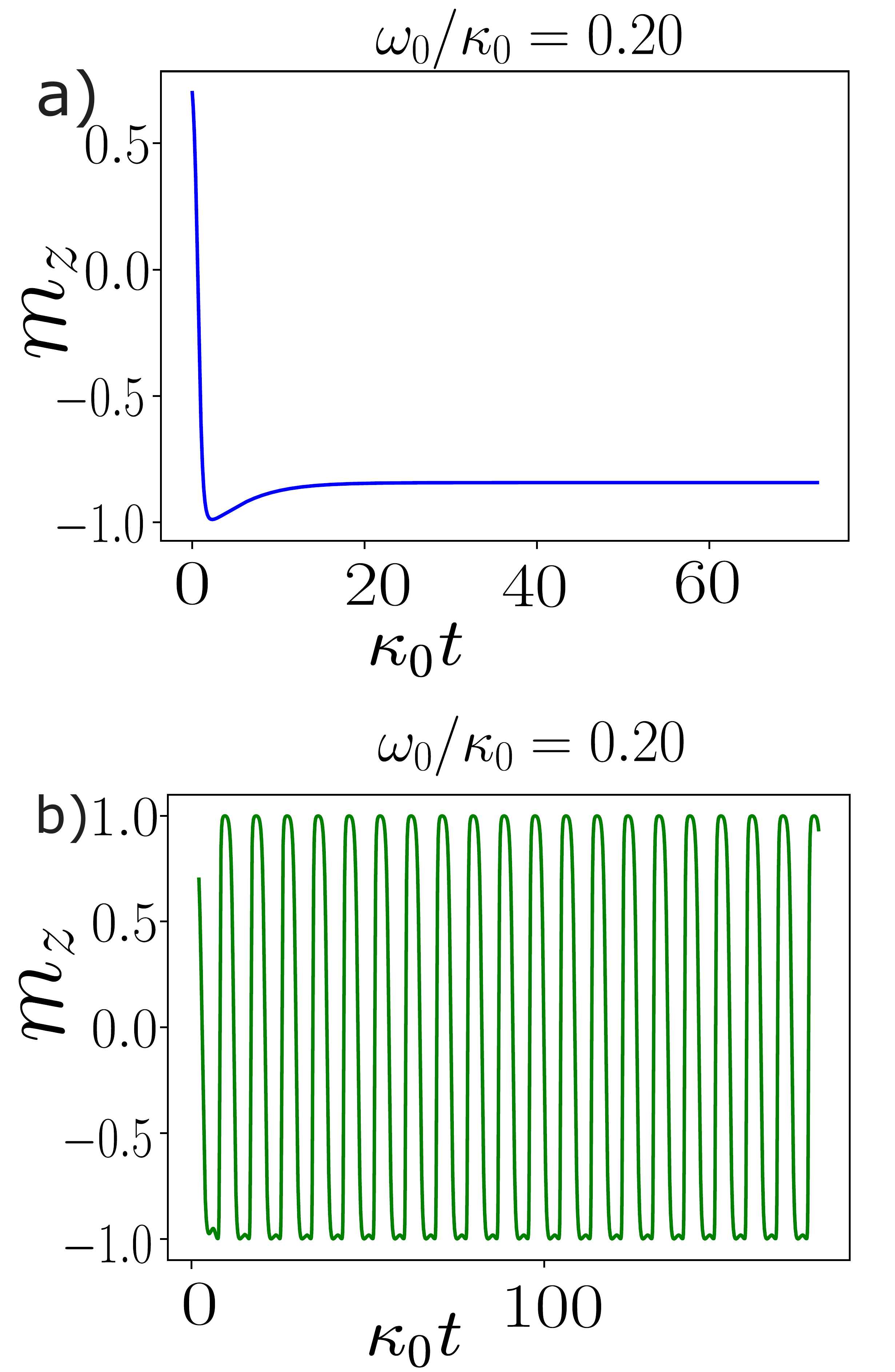}
     \caption{Time evolution of $m_z$ for $\omega_x=1,\omega_z=0.6$, in the (a) the Markovian regime with $\gamma=3\kappa_0$, and in the (b)  non-Markovian regime with modulation amplitude $\gamma = \kappa_0/4$, for $\omega_0/\kappa_0 = 0.20$. In the Markovian case, the system reaches a time-independent steady state, whereas the non-Markovian dynamics leads to a BTC phase characterized by persistent oscillations.}
     \label{fig:btcgeneral}
\end{figure}
 To examine the behavior of the BTC phase under more general conditions, we consider the case of non-zero $\omega_x$ and $\omega_z$. Our analysis reveals that in the Markovian regime characterized by strong dissipation the system relaxes to a time-independent steady state. In contrast, in the non-Markovian regime, the system exhibits persistent oscillations indicative of the BTC phase (see Fig. \ref{fig:btcgeneral}).

\section{Time-dependent decay rate}
\label{appB}

In order to motivate the phenomenological model used here with the time-dependent $\kappa(t)$ in Eq. \eqref{eqkappa},  we now present a brief derivation of this form of  $\kappa(t)$ arising in the damped Jaynes Cummings model on resonance, following Ref. \cite{breuer07the}. The damped Jaynes Cummings model comprises a two-level system coupled to a bath of harmonic oscillators, with mode-dependent coupling strengths.  In this case
 the total time-independent Hamiltonian can be written as
\begin{equation}
H = H_S + H_B + H_I ,
\end{equation}
where
\begin{equation}
H_S = \omega_S \sigma_+ \sigma_- ,
\end{equation}
and
\begin{equation}
H_B = \sum_k \omega_k b_k^\dagger b_k ,
\end{equation}
denote the system and the bath Hamiltonians, respectively, while the system-bath interaction Hamiltonian  is given by
\begin{equation}
H_I = \sigma_+ \otimes B + \sigma_- \otimes B^\dagger ,
\qquad
B = \sum_k g_k b_k .
\end{equation}
Here $\omega_S$ denotes the Rabi frequency of the two-level system, $\sigma_{\pm} = \frac{1}{2}\left(\sigma_x \pm i\sigma_y \right)$ acts on the two-level system, $b_k$ ($b_k^{\dagger}$) is the annihilation (creation) operator acting on the $k$-th mode of the bath with frequency $\omega_k$ and system-bath interaction strength $g_k$. 
Assuming the system and the bath are initially uncorrelated, and the bath is in its vacuum state at time $t = 0$, we arrive at the following time-dependent master-equation in the interaction picture \cite{hou11alternative}:
\ba
\dot{\rho} &=& -i\frac{\Omega(t)}{2}\left[\sigma_+ \sigma_-, \rho \right] \non\\ &+& \kappa(t) \left(\sigma_- \rho \sigma_+ -\frac{1}{2}\sigma_+ \sigma_-\rho  -\frac{1}{2}\rho \sigma_+ \sigma_- \right),
\ea
where $\Omega(t) = -2 \, {\rm Im}\left[\frac{\dot{c}(t)}{c(t)} \right]$ plays the role of a time-dependent Lamb shift, and $\kappa(t) = -2 \, {\rm Re}\left[\frac{\dot{c}(t)}{c(t)} \right]$ denotes a time-dependent decay rate. Here $c(t)$ satisfies the equation
$\dot{c}(t) = -\int_0^t \Phi(t - t^{\prime}) c(t^{\prime}) dt^{\prime}$. The bath-correlation function for times $t$ and $t^{\prime}$ is given by 
\ba 
\Phi(t - t^{\prime}) &=& {\rm Tr}_B \{B(t)B(t^{\prime}) \} e^{i \omega_S(t-t^{\prime})} \non\\
&=& \int d\nu J(\nu) e^{i (\omega_S - \nu)(t-t^{\prime})}.
\ea
Assuming a Lorentzian bath spectral function 
\ba
J(\nu) = \frac{1}{2\pi}\frac{\kappa_0 \gamma^2}{\left(\omega_S - \nu \right)^2 + \gamma^2},
\ea
and restricting ourselves to the case of a single excitation in the atom cavity system, we get $\Omega(t) = 0$ and $\kappa(t)$ given by Eq. \eqref{eqkappa} with $\kappa_{max} \to \infty$. Consequently we arrive at a master equation with a time-dependent decay rate, arising in the presence of a time-independent Hamiltonian.

\begin{widetext}

\section{Robustness against initial states}
\label{appC}

In order to test the robustness of the dynamics with respect to the choice of initial state, we consider several arbitrary initial states in the non-BTC, BTC and HO-LC phases.  As shown in Fig. \ref{fig:initialcondition},
all trajectories converge   in the BTC and HO-LC phases (Figs. \ref{fig:initialcondition}b and \ref{fig:initialcondition}c), while the FFT peaks overlap (Figs. \ref{fig:initialcondition}e and \ref{fig:initialcondition}f), thus emphasizing the robustness of these phases to the choice of the initial state. On the other hand, the long time dynamics and the FFT depends on the choice of the initial state in the non-BTC phase (Figs. \ref{fig:initialcondition}a and \ref{fig:initialcondition}d).

\begin{figure}[H]
    \centering
     \includegraphics[width=\textwidth]{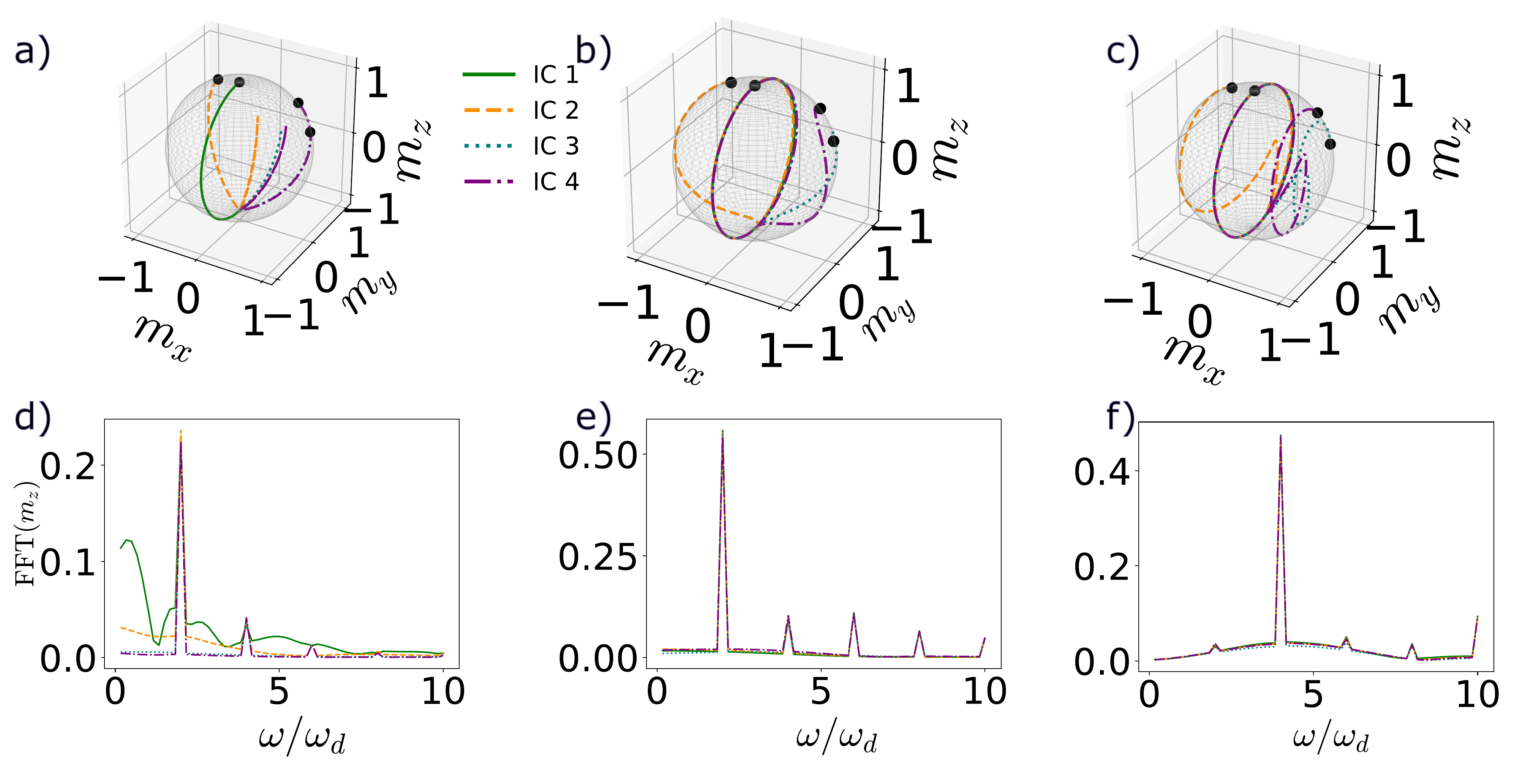}
    \caption{
    Bloch-sphere trajectories for 4 different initial states, labeled (IC(1-4)), 
    (a) for the non-BTC phase at $\omega_0/\kappa_0 = 0.02$, 
    (b) BTC phase at $\omega_0/\kappa_0 = 0.4$, and
    (c) HO-LC phase $\omega_0/\kappa_0 = 1.4$. The corresponding FFTs are plotted in (d), (e) and (f), respectively.
    The initial sates are
    $(m_x^0,m_y^0,m_z^0) = {\rm IC}1:~(0,0,1), {\rm IC}2:~(-0.58,0.42,0.70), {\rm IC}3:~(0.70,0.70,0.10),  {\rm IC}4:~(\frac{1}{\sqrt{3}},\frac{1}{\sqrt{3}},\frac{1}{\sqrt{3}}     )$. As shown in (a) and (d), the final state as well as the corresponding FFT are highly sensitive to the choice of the initial state in the non-BTC phase for small $\omega_0/\kappa_0$, while the final limit cycle as well as the FFT peaks are robust to the choice of the initial state in the (b) and (e) BTC and (c) and (f) HO-LC phases. Here $\gamma/\kappa_0 = 1/4$.
  }
    \label{fig:initialcondition}
\end{figure}
\section{Non-BTC regime for very low frequency}
\label{appD}

For $\frac{\omega_0}{\kappa_0} \ll 1$, the system exhibits periodic oscillations in the non-Markovian regime. We note that this phase is distinct from the BTC phase, as   evident from the Bloch sphere representation, which shows that these oscillations do not correspond to a limit cycle  (see Fig.~\ref{fig:smallomega}). Furthermore, as shown in Fig. \ref{fig:initialcondition} a, the long time dynamics depends on the choice of the initial state in this phase. 

\begin{figure}[H]
     \centering
     \includegraphics[width=\linewidth]{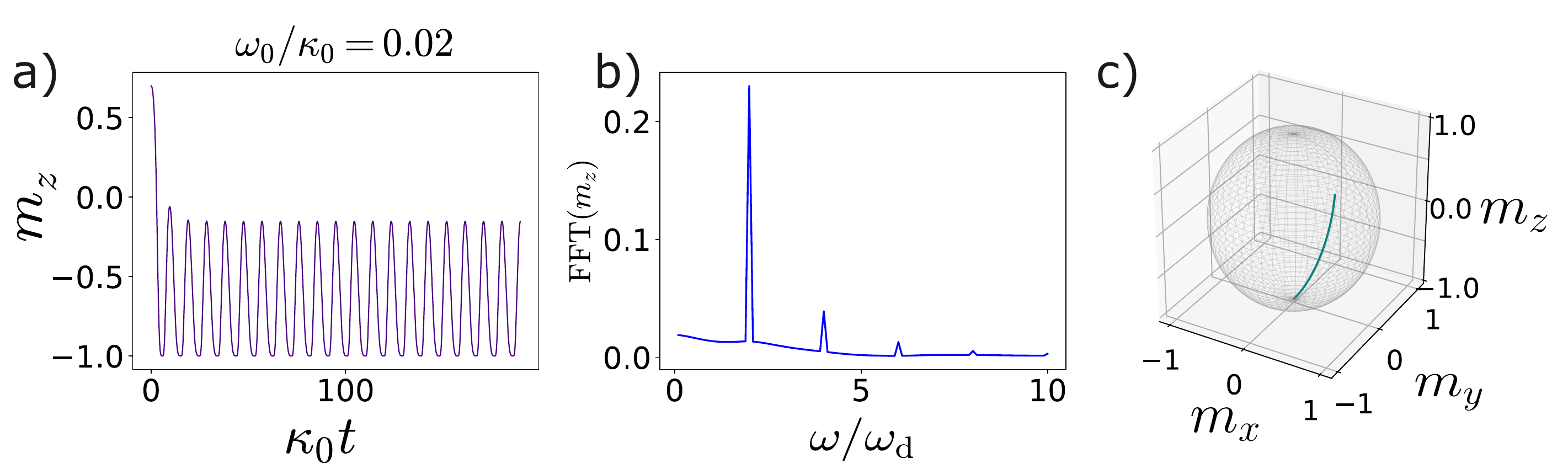}
      \caption{Figure showing (a) $m_z$ as a function of time,  the corresponding (b) FFT and  (c) the Bloch sphere representation plotted for the last 10 cycles of the BTC, in the non-Markovian regime, for $\omega_0/\kappa_0=0.02$. As shown by the Bloch sphere representation, this dynamics does not correspond to a limit cycle, and hence cannot be deemed as a BTC. Here  $\gamma=\kappa_0/4$.}

     \label{fig:smallomega}
 \end{figure}

\end{widetext}

% \bibliography{refs_TC}
% \bibliographystyle{unsrt}
\end{document}